\documentclass[
    aip,
    jcp,
    reprint,
    amsmath,
    amssymb,
    superscriptaddress,
    floatfix
]{revtex4-2}

\usepackage[english]{babel}
\usepackage{amsmath,amssymb,amsfonts,bm,mathtools}
\usepackage{graphicx}
\usepackage{xcolor}
\usepackage{siunitx}
\usepackage[version=4]{mhchem}
\usepackage{orcidlink}

\usepackage{hyperref}

\usepackage[capitalise]{cleveref}

\newcommand{\bl}{\pmb{\lambda}}
\newcommand{\bchi}{\pmb{\chi}}

\begin{document}

\title{Polarizable Embedding for Photoelectron Spectroscopy and Photoelectron Circular Dichroism in Solution:
Core-Level Ionization of Aqueous Alanine}

\author{Giovanni Nottoli\,\orcidlink{0000-0002-5310-8299}}
\affiliation{Scuola Normale Superiore, Piazza dei Cavalieri 7, 56126 Pisa, Italy}

\author{Piero Decleva\,\orcidlink{0000-0002-7322-887X}}
\affiliation{Universit\`a degli Studi di Trieste, Via Licio Giorgieri 1, 34127 Trieste, Italy}

\author{Chiara Cappelli\,\orcidlink{0000-0002-4872-4505}}
\email{chiara.cappelli@sns.it}
\affiliation{Scuola Normale Superiore, Piazza dei Cavalieri 7, 56126 Pisa, Italy}

\begin{abstract}
We present a polarizable quantum-mechanics/molecular-mechanics formulation for the calculation of molecular photoionization observables in solution. The approach couples the static-exchange density-functional-theory treatment of bound and continuum electronic states with a fully atomistic polarizable embedding described with the fluctuating-charge force-field. The fluctuating charges are determined self-consistently with the ground-state quantum-mechanical density and are subsequently included in the static-exchange Hamiltonian to account for the outgoing electron. The numerical behavior of the coupled scheme is validated and applied to the X-ray photoelectron spectrum and photoelectron circular dichroism of zwitterionic L-alanine in aqueous solution, using an ensemble of configurations extracted from molecular dynamics. Calculated spectra reproduce experimental profiles remarkably well, demonstrating that atomistic solvent structure and environmental polarization must be treated together to reliably model photoelectron observables in aqueous solution.
\end{abstract}

\keywords{photoionization in solution, photoelectron circular dichroism, polarizable embedding,
fluctuating charges, continuum electronic states, aqueous alanine}

\maketitle

\section{Introduction}

Photoionization and related photoelectron processes play a central role in modern physical chemistry, providing direct access to the electronic structure and dynamics of molecular systems. Their relevance has further increased in recent decades with the continuous development of synchrotron radiation sources and, more recently, ultrafast XUV and X-ray light sources, which provide intense, tunable, and polarization-controlled radiation over broad spectral ranges. Photoelectron spectroscopy and related techniques offer an exceptionally detailed view of electronic structure and electronic dynamics, combining energy, site, and symmetry selectivity with temporal resolution that is difficult to achieve with other experimental approaches. \cite{suzuki2019ultrafast,jordan2020attosecond,woerner2025ultrafast}

While photoelectron spectroscopy of gas-phase molecules and solid samples developed rapidly after the introduction of the technique, its application to liquids proved considerably more difficult and remained largely dormant for many years after the first unsuccessful attempts. The field was revived by the development of dedicated liquid-microjet sources and their combination with modern synchrotron facilities, \cite{winter2006photoemission} opening direct access to the electronic structure of liquid water and aqueous solutions. Since then, liquid-phase photoelectron spectroscopy has developed into a powerful approach for investigating solvation, electronic structure, and chemical processes in aqueous environments. \cite{seidel2016valence,suzuki2019ultrafast} This development is particularly significant considering that solutions, and aqueous solutions in particular, are at the heart of chemistry and play a central role in the life sciences.

The development of theoretical models for the description and interpretation of these experiments has lagged further behind. While theory has been essential for the advancement and deeper understanding of gas-phase photoionization,\cite{fransson2013carbon,
wenzel2014calculating,wenzel2015analysis,brabec2012communication,
sen2013study,peng2015energy,coriani2015comm,
vidal2019fcCVSEOM,vidal2020equation,
park2022mrtddft,norman2018simulating,bokarev2020theoretical} theoretical studies of molecules in liquids have so far mainly focused on photoelectron binding energies, spectral shifts, and related energetic properties. \cite{winter2005electron,yepes2014photoemission} A molecular-level description of genuine continuum observables, such as photoionization cross sections and photoelectron angular distributions, is considerably more demanding and remains much less developed. This limitation becomes particularly important for observables whose magnitude and sign depend directly on the detailed structure of the continuum wavefunction. To the best of our knowledge, photoelectron circular dichroism (PECD) has not yet been theoretically simulated for a molecular system in solution. In this work, we therefore move beyond the calculation of liquid-phase photoelectron binding energies and spectra and address a genuinely continuum-sensitive chiroptical observable in aqueous solution.

Photoelectron circular dichroism is the forward--backward asymmetry in the emission of photoelectrons from randomly oriented chiral systems irradiated with circularly polarized light. \cite{turchini2004circular,stener2004density,waniepecd} Unlike conventional electronic circular dichroism, PECD arises within the electric-dipole approximation and can reach comparatively large magnitudes. Its sign and intensity are determined by the interference of continuum partial waves and are therefore highly sensitive to molecular geometry, the emitting site, and the electronic potential sampled by the outgoing electron. Most theoretical treatments of PECD have so far focused on isolated molecules or small clusters. \cite{powis2000photoelectron,ritchie1976theory,decleva2022continuum,nahon2015valence}

Extending such continuum descriptions to aqueous solutions poses additional and qualitatively different challenges. As emphasized by Signorell and Winter,\cite{signorell2022photoionization} the retrieval and interpretation of photoelectron angular information from aqueous systems are intrinsically complicated by the interaction of the emitted electron with the surrounding liquid, including elastic and inelastic scattering during its transport toward the liquid--vacuum interface. \cite{signorell2022photoionization} At the molecular level, however, there is an additional and complementary problem: even before such transport effects are considered, the photoelectron continuum is formed in the microscopic potential generated by the surrounding solvent. A theoretical treatment of photoionization in solution must therefore account for the effect of the environment not only on the initial bound electronic state, but also on the continuum wavefunction itself.

These effects arise at two closely related levels. First, the electronic structure of the solute is modified by the solvent through electrostatic interactions, mutual solute-solvent polarization and, depending on the chemical nature of the solute, specific interactions such as hydrogen bonding. Second, the continuum wavefunction of the outgoing electron is generated in the anisotropic potential of the molecular environment and can therefore be directly affected by the instantaneous solvent configuration. Continuum dielectric approaches \cite{tomasi2005quantum,Klamt1993,Klamt1995} provide a possible route to account for solvent polarization at an average level, but do not retain the molecular structure of the solvent.\cite{giovannini2023continuum} Conversely, fully atomistic fixed-charge embedding preserves the microscopic description of the environment, but neglects mutual solute-solvent polarization.\cite{giovannini2020molecular} A molecular description of photoionization in solution thus calls for an embedding framework capable of treating both the microscopic structure and the electronic response of the environment, while remaining sufficiently efficient to allow configurational sampling.\cite{giovannini2020molecular}

In this work, we introduce the first quantum-mechanical (QM)/ Molecular Mechanics (MM)-like polarizable embedding formulation for molecular photoionization in the continuum. The approach couples the static-exchange density-functional-theory (SE-DFT) framework \cite{decleva2022continuum,toffoli2024tiresia} with the fully atomistic, polarizable fluctuating-charge (FQ) model. \cite{cappelli2016integrated,giovannini2020molecular} The ground-state density of the solvated solute and the FQ charges are first determined self-consistently, thus accounting for mutual solute-solvent polarization. The resulting FQ charges are then explicitly included in the static-exchange Hamiltonian used to construct the continuum states. As a consequence, both the bound electronic structure and the outgoing photoelectron experience the atomistically resolved, polarizable solvent environment. The method therefore extends polarizable embedding from the description of bound-state and energetic properties to genuine photoionization continuum observables, while retaining a numerical efficiency compatible with extensive configurational sampling of the solvated system.

We apply this framework to zwitterionic L-alanine in aqueous solution. Moving beyond the calculation of photoelectron binding energies and X-ray photoelectron spectra, we exploit the explicit description of the continuum to simulate the carboxylate-carbon PECD, providing, to the best of our knowledge, the first theoretical PECD spectrum of a molecule in solution, and specifically in water. We consider two complementary observables, namely the C~$1s$ photoelectron spectrum and the carboxylate-carbon PECD, and compare nonpolarizable and polarizable descriptions of the environment. The photoelectron spectrum primarily probes the local, site-dependent solvent screening of the three chemically distinct carbon atoms, whereas PECD provides a considerably more stringent test of the continuum description, since it depends on the phases and interference of the outgoing partial waves and is therefore directly sensitive to the microscopic potential generated by the solvent. The photoionization calculations are furthermore combined with extensive configurational sampling extracted from molecular dynamics simulations. \cite{giovannini2020molecular,gomez2023multiple}

Recent liquid-jet measurements by Stemer \textit{et al.} \cite{stemer2025photoelectron} on aqueous alanine provide a particularly demanding experimental reference for this purpose. Their study established core-level PECD for the three protonation states of alanine and demonstrated that the chiral photoelectron response can retain site-specific and structure-sensitive information in the aqueous environment. For zwitterionic alanine at approximately neutral pH, the measured carboxylate-carbon PECD is weak and structured, with substantial uncertainties over part of the available kinetic-energy range. This dataset nevertheless provides a particularly valuable benchmark for the present approach, because the corresponding C~$1s$ photoelectron spectrum independently probes the site-specific solvent response at the three chemically distinct carbon atoms. Taken together, the two observables therefore allow us to assess complementary aspects of the embedding description: the effect of solute-solvent polarization on core-level energetics and the sensitivity of the outgoing photoelectron continuum, and hence of PECD, to the microscopic solvent environment.

\section{Theoretical Methodology}

This section presents the theoretical framework of our methodology. We begin with a brief review of the fully polarizable QM/Fluctuating Charge (FQ) approach. We then focus on the development of our method, which formulates a novel Static-Exchange SE-DFT Hamiltonian for the embedded FQ system. Within this integrated framework, electrostatic, polarization, and hydrogen bonding solvent contributions are fully and rigorously recovered in the continuum state calculations through the direct, self-consistent incorporation of the FQ terms into the photoelectron scattering Hamiltonian. In the following, in line with the SE-DFT framework, we focus on a DFT description of the QM portion of the system (the \textit{solute}).

\subsection{Polarizable QM/FQ Approach}

In the polarizable QM/FQ framework, the system is treated at the full atomistic level, and is partitioned in a QM portion (the \textit{solute}) and a classical environment (the \textit{solvent}). The solvent, which is atomistically described, is modelled by means of the fully atomistic, polarizable FQ force field \cite{giovannini2020molecular, cappelli2016integrated}. Each atom belonging to molecules of the classical FQ portion is assigned an atomic charge $q$. FQ charges $\mathbf{q}$ are not fixed, but vary so that their values satisfy the Electronegativity Equalization Principle (EEP). \cite{mortier1985electronegativity, rick1994dynamical} When coupled to a QM description of a region of the system (the \textit{solute} in this specific case) the FQ charges are adjusted to the QM region, so that the charge distribution reaches full equilibrium with the electrostatic potential generated by both the classical solvent charges and the QM density. FQ charges are parametrized in terms of atomic electronegativity and chemical hardness, which constitute the sole parameters of the model. \cite{rick1996dynamical, giovannini2016effective}

In the hybrid QM/FQ approach, the total energy of the system reads as follows:
\begin{equation}
E_{\text{tot}} = E_{\text{QM}}[\mathbf{P}] + E_{FQ}(\mathbf{q}, \bl) + E_{int}(\mathbf{P}, \mathbf{q})
\label{eqtot}
\end{equation}
where $\mathbf{P}$ is the QM density matrix and $\mathbf{q}$ represents the vector of classical fluctuating charges \cite{cappelli2016integrated, giovannini2020molecular}. 
$E_{\text{FQ}}$ is defined as:
\begin{equation}
E_{\text{FQ}}(\mathbf{q}, \bl) = \mathbf{q}^\dagger \bchi + \frac{1}{2} \mathbf{q}^\dagger \mathbf{T}_{qq} \mathbf{q} - \bl^\dagger (\mathbf{Q} - \mathbf{1}_\lambda \mathbf{q})
\end{equation}
Here, $\bchi$ and $\mathbf{T}_{qq}$ denote atomic electronegativities and the charge-charge interaction tensor, respectively \cite{cappelli2016integrated}. To prevent the so-called \textit{polarization catastrophe} at short distances, the interaction tensor elements are typically modeled using the Ohno kernel:\cite{cappelli2016integrated, giovannini2020molecular}
\begin{equation}
J_{ii} = \eta_i, \quad J_{ij} = \frac{1}{\sqrt{r_{ij}^2 + (\frac{2}{\eta_i + \eta_j})^2}}
\end{equation}
where $\eta_i$ represents the chemical hardness. The Lagrangian multipliers $\bl$ enforce the conservation of the total molecular charge $\mathbf{Q}$ on the solute molecule. \cite{cappelli2016integrated}

The QM/FQ coupling is defined by the electrostatic interaction between the QM electronic density and the classical fluctuating charges:\cite{cappelli2016integrated, giovannini2020molecular}
\begin{equation}
E_{int} = \mathbf{q}^\dagger \mathbf{V}(\mathbf{P})
\end{equation}
where $\mathbf{V}(\mathbf{P})$ is the electrostatic potential generated by the QM density at the sites of the classical FQ atoms. \cite{cappelli2016integrated}. 

\subsection{Polarizable QM/FQ embedding within a static-exchange Hamiltonian}

In SE-DFT, the electronic continuum is described by solving the one-electron Schrödinger equation within a finite spherical domain of radius $R_{\text{max}}$ using a multicenter basis set of B-splines coupled with real spherical harmonics \cite{decleva2022continuum, toffoli2024tiresia}:
\begin{equation}
\chi_{ilm}(r, \theta, \phi) = \frac{1}{r} B_i(r) Y_{lm}(\theta, \phi)
\end{equation}
The basis set consists of a long-range One-Center Expansion (OCE) to describe the oscillatory behavior of the continuum and short-range multicenter functions (LCAO) centered on nuclei to represent Coulomb cusps \cite{decleva2022continuum, toffoli2024tiresia}. A fixed Hamiltonian ($H_{\text{SE-DFT/FQ}}$) is constructed using the ground-state density $\rho$ generated from a separate DFT calculation. \cite{decleva2022continuum, toffoli2024tiresia}

QM/FQ and SE-DFT are integrated by including the contribution of the FQ charges into the SE-DFT Hamiltonian. This corresponds to 
adding an external potential to the Kohn-Sham (KS) Hamiltonian, arising from the FQ charges computed at convergence at the end of the ground-state KS/FQ calculation:

\begin{align}
    H_{\text{SE-DFT/FQ}} = H_{\text{SE-DFT}} + V_{\text{FQ}}(q) =  \nonumber \\ 
    -\frac{1}{2} \nabla^2 + V_{en} + V_C(\rho) + V_{XC}(\rho) + \sum_p \frac{q_p}{|\mathbf{r} - \mathbf{r}_p|}
    \label{hamilt}
\end{align}

The last term in eq.\ref{hamilt} represents the external potential produced by the FQ charges, which polarizes the QM density. 

In the following discussion, results denoted as FQ are obtained with the parameterization developed in Ref.\cite{giovannini2019effective}

For the sake of comparison, we shall also present some additional results obtained with alternative solvent models, i.e.
\begin{enumerate}
    \item TIP3P, which is a non polarizable (fixed charges) electrostatic embedding scheme; \cite{mark2001structure}
    \item  $\text{FQ}_{\text{R}}$, which keeps the FQ framework but employs a different parametrization, namely that taken from Ref. \citenum{rick1994dynamical} ;
    \item COSMO where the solvent is described implicitly as a polarizable continuum, so no explicit statistical sampling is involved.  \cite{Klamt1993}
\end{enumerate}

\section{Computational Details}
\label{sec:computational_details}

Zwitterionic L-alanine in aqueous solution was chosen as the target chiral system. Molecular configurations were extracted from classical molecular dynamics (MD) trajectories \cite{prusa2020dataset}, from which representative snapshots were selected for ensemble averaging.

For each snapshot, a spherical solvent droplet was generated by cutting the explicit water box around the zwitterionic solute. 
To guarantee convergence, the droplet radius was tested between $10$~\AA~ and $30$~\AA. 

To systematically evaluate the impact of environmental polarization, in addition to the present FQ model, we compare our results with two atomistic solvent models, the non-polarizable TIP3P, and the alternative parameterization $\text{FQ}_{\mathrm{R}}$, and the implicit continuum scheme COSMO  with target molecular geometries optimized at the LB94/TZ2P level. Reference calculations were also performed in the gas phase (vacuum).

The computational workflow integrates two distinct software engines:
\begin{enumerate}
    \item \textbf{Amsterdam Modeling Suite (AMS / ADF engine):\cite{baerends2025amsterdam}} Used to compute the reference ground-state electronic densities, KS-DFT bound-state wavefunctions, and initial molecular orbital energies, including the $\text{FQ}$ charges self-consistently.
    \item \textbf{\textit{Tiresia} Code:\cite{toffoli2024tiresia}} Employed to calculate the continuum photoelectron wavefunctions, partial photoionization cross sections ($\sigma$), and photoelectron circular dichroism ($\beta_1$ parameters) \cite{toffoli2024tiresia}. \textit{Tiresia} utilizes a B-spline One-Center Expansion (OCE) scheme that explicitly incorporates the electrostatic embedding potential generated by the classical solvent environment (fixed charges or fluctuating charges) into the continuum Hamiltonian.
\end{enumerate}

The converged ADF density and FQ charges were transferred to
\textit{Tiresia}. Continuum states were expanded in the combined OCE/LCAO
B-spline basis. The convergence tests described below were used to select
$L_{\max}=20$ and an OCE matching radius extending by at least
approximately $10$~\AA\ beyond the explicit solvent sphere. X-ray photoelectron spectroscopy (XPS) spectra of L-alanine were generated by exploiting 200 spherical droplets with radius equal to 
$20$~\AA\, and $R_{\text{max}} = 60~\text{a.u.}$ ($31.75$~\AA). A linear radial grid with spacing of $0.5$ a.u. was used, generating 143 B-splines, with a total basis size, including the LCAO functions, of 63765 for this case. In the case of PECD, achieving convergence on a threshold of 0.15\% required around 600 structures. The plots reported in the following sections refer to averaging over 1000 spherical droplets.

For XPS photoelectron cross-sections, the unbound continuum electronic states were evaluated within a kinetic energy window tailored to the experimental conditions. Considering the experimental monochromatic photon energy of $h\nu = 480.46\,\text{eV}$, the photoelectron kinetic energy ($E_k$) for each molecular snapshot was determined by subtracting the corresponding calculated core binding energy from $h\nu$. Finally, a narrow kinetic energy interval of $\pm 0.002\,\text{a.u.}$ centered around each snapshot-specific $E_k$ value was sampled, and the median intensity across this window was extracted to generate the final spectral profiles.

To enable a direct and refined comparison with low-energy experimental PECD data, broad exploratory continuum calculations were performed over a photoelectron kinetic energy range from $0$ to $2$ Hartree (approximately $0$--$54.4\,\text{eV}$) sampled across 130 energy points for 200 snapshots.  A non-uniform continuum grid was adopted: the range from $0$ to $0.6\,\text{Hartree}$ was discretized in steps of $0.01\,\text{Hartree}$, whereas the higher-energy region from $0.6$ to $2.0\,\text{Hartree}$ was sampled with a step size of $0.02\,\text{Hartree}$.\cite{decleva2022continuum,toffoli2024tiresia}.

Subsequently, to directly compare and match our calculations with the experimental results, a targeted analysis was conducted on the 10 energy points reported experimentally~\cite{stemer2025photoelectron}. For these specific points, the ensemble size was expanded to 1000 snapshots in order to minimize the statistical uncertainty.

\begin{figure}[t]
    \centering
    \includegraphics[width=0.9\linewidth]{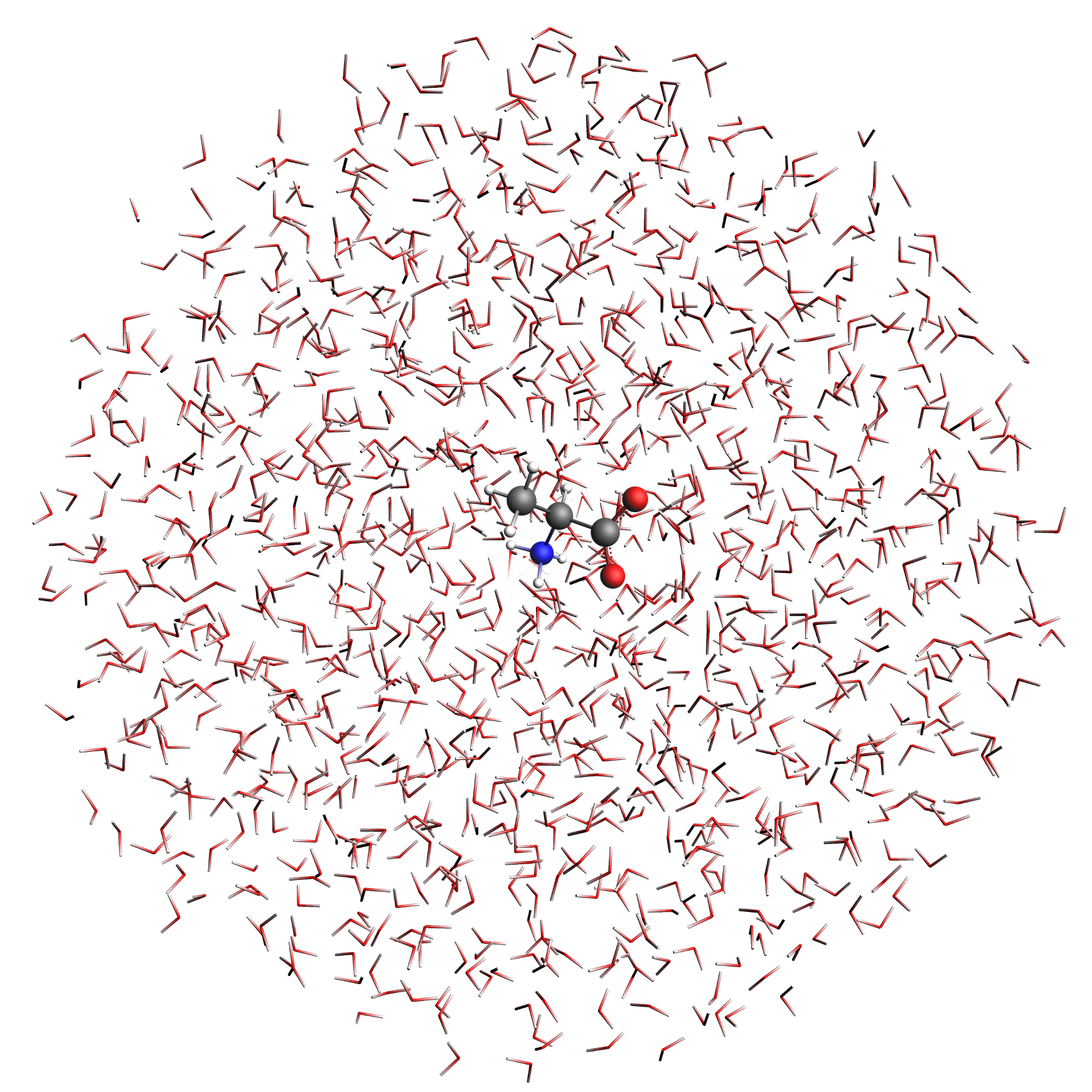}
    \caption{Representative droplet used in SE-DFT/FQ
    calculations. Zwitterionic L-alanine is located at the center of a
    spherical region of classical FQ water molecules. radius=20\AA}
    \label{fig:droplet}
\end{figure}

\section{Results and Discussion}

To benchmark the model against experimental observations, we focus our investigation on zwitterionic L-alanine in aqueous solution.\cite{stemer2025photoelectron} This system serves as an ideal prototype, as it has been widely characterized in the literature through both PECD and core-level XPS measurements. \cite{stemer2025photoelectron} Properly reproducing these experimental spectra in the condensed phase is extremely challenging, as it requires a balanced description of both short-range hydrogen bonding interactions and long-range solvent polarization effects.

To ensure a highly accurate and numerically stable application of the novel approach, a rigorous assessment of the technical parameters required by the computational framework is mandatory. Specifically, we systematically investigate the convergence and stability of the photoionization calculations with respect to three key factors: the matching radius of the B-spline grid ($R_{\text{max}}$), the maximum angular momentum of the one-center expansion ($L_{\text{max}}$), and the radius of the classical FQ solvation shell. In addition, the quality of the classical FQ embedding is benchmarked with respect to full QM values for a prototypical system.

\subsection{Method validation and numerical convergence}

To validate the performance of SE-DFT/FQ against full QM reference calculations, we focus on a microsolvation model. Specifically, the water dimer is selected to examine the recovery of the target electronic properties and to appreciate how the FQ approach performs compared to a full QM description.
SE-DFT/FQ reproduces the electrostatic
and polarization effects of the full-QM reference with remarkable accuracy, but at a substantially lower
computational cost. 

We move on to validating the approach with respect to the parameters specific to the
continuum-state expansion. To this end, we resort to the representative spherical droplet shown in Fig.\ref{fig:droplet}, and systematically probe several critical computational parameters: OCE radius $R_{\text{max}}$ and maximum angular momentum $L_{\text{max}}$. While such parameters have been extensively investigated and are well-established for isolated systems in the gas phase, their behavior and convergence properties require a deeper understanding when moving to condensed phases treated by means of SE-DFT/FQ. Finally, the model is validated with respect to the radius of the FQ solvation shell and
different choices for the single-center expansion origin (i.e., the central atom of the solute molecule).

\paragraph{Radial Cut-off of the One-Center Expansion ($R_{\text{max}}$)}

The spatial extent of the radial grid ($R_{\text{max}}$) is probed for a given solvation sphere to guarantee that the whole explicit potential is properly enclosed and reconnected with the asymptotic continuum wavefunctions.

To accurately describe the electronic continuum, the radius of the one-center expansion (OCE) must be sufficiently large to encompass the entire molecular potential, including the explicit solvation shell. In the \textit{Tiresia} framework, the numerical solution of the one-particle Schrödinger equation within the spherical box is matched at $R_{\text{max}}$ to the analytical asymptotic form (Coulomb or plane waves), ensuring continuity of the wavefunction and its first derivative. To validate the required radial extent, we performed a series of benchmark calculations on a single configurational snapshot by systematically varying the OCE radius. By construction, $R_{\text{max}}$ must exceed the radius of the explicit water sphere to ensure that the potential has reached its monopolar limit. Numerical results indicate that convergence of the photoionization cross section $\sigma$ is rapidly achieved, becoming stable when the radius extends at least $10$ \AA{} beyond the solvation shell. 
Consequently, this margin was adopted for subsequent production runs to optimize computational efficiency without sacrificing accuracy.

\paragraph{Maximum Angular Momentum of the OCE ($L_{\text{max}}$)}

The OCE basis set requires a high angular momentum expansion to accurately describe the pronounced anisotropy of the molecular potential and the contribution of high-$L$ partial waves in the electronic continuum. Given that the presence of an explicit aqueous environment significantly breaks the molecular symmetry and extends the spatial range of the potential, we systematically investigated the sensitivity of the photoionization observables to the maximum angular momentum parameter, $L_{\text{max}}$. Simulations were carried out with $L_{\text{max}}$ values ranging from $16$ to $24$, while fixing the matching radius ($R_{\text{max}}$) at the previously optimized value of $31.75$ \AA{} ($60\,\text{a.u.}$).

The quality of the expanded basis set was initially assessed by comparing the bound-state eigenvalues computed by \textit{Tiresia} with the reference values obtained from the preliminary ground-state \textit{ADF} calculation. An agreement within a threshold of $0.15\,\text{eV}$ for the valence orbitals was deemed sufficient to confirm that the basis set accurately captures the local potential, ensuring a reliable description of the corresponding continuum states. This stringent condition is fully satisfied when choosing $L_{\text{max}} = 20$.

Subsequently, the photoionization observables were monitored as a function of the angular momentum cutoff. Within the $18$--$21$ range, the variations in the computed properties with respect to changes in $L_{\text{max}}$ were found to be marginal, demonstrating that the properties are already well converged at $L_{\text{max}} = 18$. Nevertheless, to guarantee the highest possible accuracy for both the bound eigenvalues and the continuum wavefunctions, consistent with the optimization discussion in the previous section, a value of $L_{\text{max}} = 20$ was selected for all subsequent production calculations.

\paragraph{Radius of the FQ Solvation Shell}

To evaluate the impact of the FQ solvation shell size on the computed photoionization properties, spherical droplets of varying radii, ranging from $10\,\text{\AA}$ to $30\,\text{\AA}$, were studied. The largest radius corresponds to the bulk-like solvation limit (encompassing roughly 8,000 water molecules at a $30\,\text{\AA}$ radius). Our analysis indicates that the statistical convergence of the photoionization observables is robustly achieved at a radius of approximately $16\,\text{\AA}$. A conservative radius of $20\,\text{\AA}$ was selected for the explicit water shell in all production calculations. At this cut-off distance, the number of explicitly included solvent molecules is of the order of $1,000$, with minor fluctuations depending on the specific droplet.

\paragraph{Choice of the center of the OCE}

\begin{figure}[htbp]
    \centering
    \includegraphics[width=0.7\linewidth]{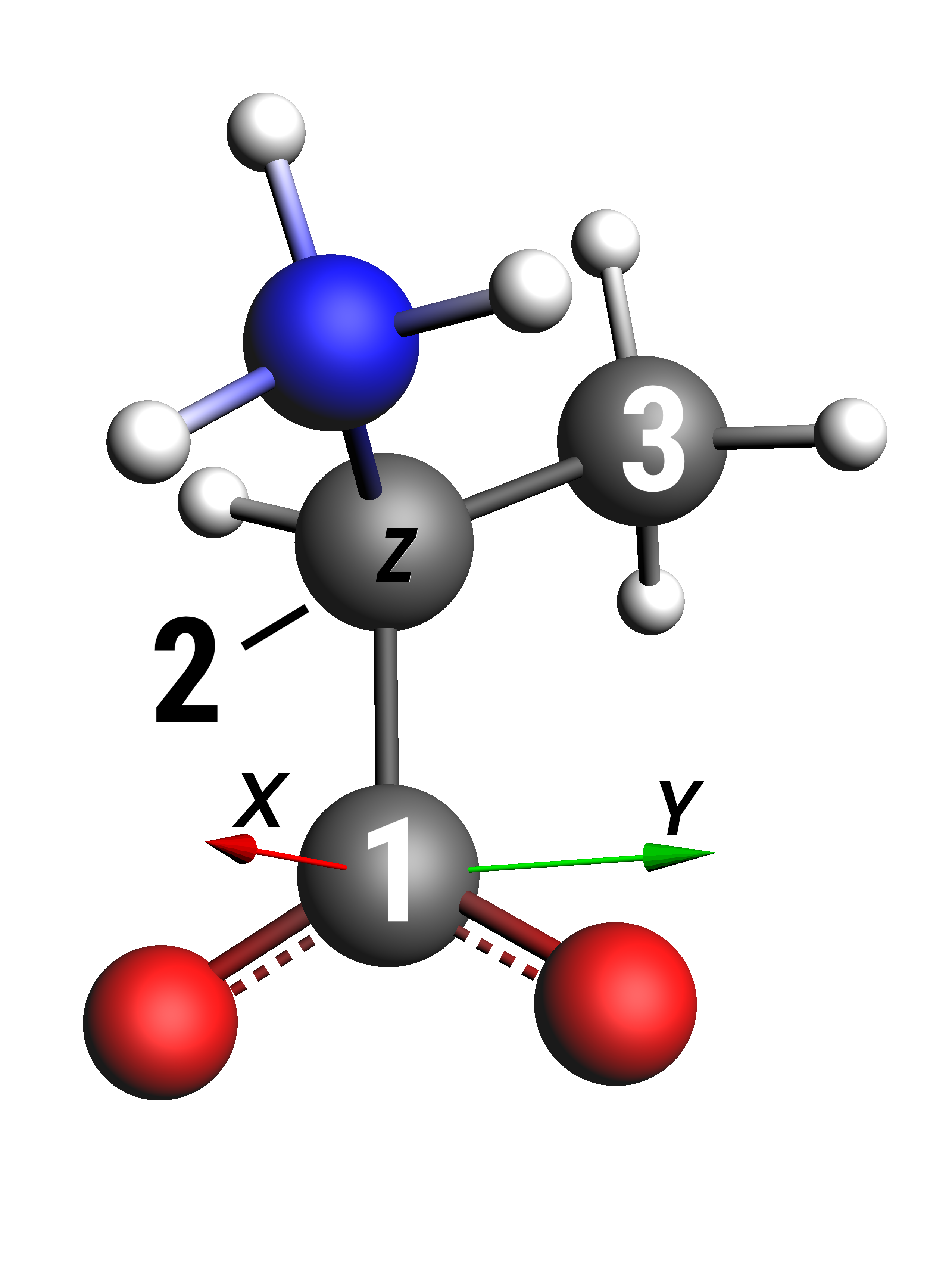}
        \caption{Molecular structure and atom numbering scheme of L-alanine in its zwitterionic form. Color code: carbon (grey), nitrogen (blue), oxygen (red), and hydrogen (white). The local reference Cartesian frame ($x, y, z$) centered on the carboxylic carbon atom (C$_1$) is shown.}
    \label{fig:alanine_number}
\end{figure}

Although fully converged results are formally independent of the choice of origin of the OCE, the coordinate origin may impact the convergence rate significantly. This sensitivity arises because the multi-center molecular potential and wavefunctions must be projected onto a single-center basis set of spherical harmonics and radial functions. Since atoms located far from the expansion center slow down convergence, placing the origin near the geometric center of the system is typically optimal. In solvated systems, the surrounding solvent environment also plays a role in determining this position. Based on convergence tests, placing the OCE origin on the carboxylic carbon of L-alanine (see Fig.~\ref{fig:alanine_number} for atom labeling) provides an optimal compromise and is maintained across all sampled geometries.

\subsection{Photoelectron spectra and PECD of aqueous L-Alanine}

\subsubsection{X-ray photoelectron spectroscopy (XPS) spectra}

The calculation of ionization potentials (IP) is a pure bound state problem, which does not involve the continuum, although the latter enters in the evaluation of band intensities and shapes. This is an important issue, which has been very extensively investigated for gas phase ionization, \cite{winter2005electron,yepes2014photoemission,ghosh2012firstprinciple} but only sparsely for solutions.

Apart from the QM approach employed, what is new in the solution case is the effect of the solvent on the IPs. This is tied to the modification of the solute ground state electronic structure, and also by the additional response of the solvent upon ionization. Intuitively, one expects an additional relaxation effect, and screening of the charge, which should decrease the IP. Here we shall limit ourselves to the ground state effect on the alanine C1s ionizations, describing the IPs at the Koopmans level, with attention to the polarization effect of the solvent and the difference in the solvation models.

\begin{figure}[t]
    \centering
    \includegraphics[width=0.30\linewidth]{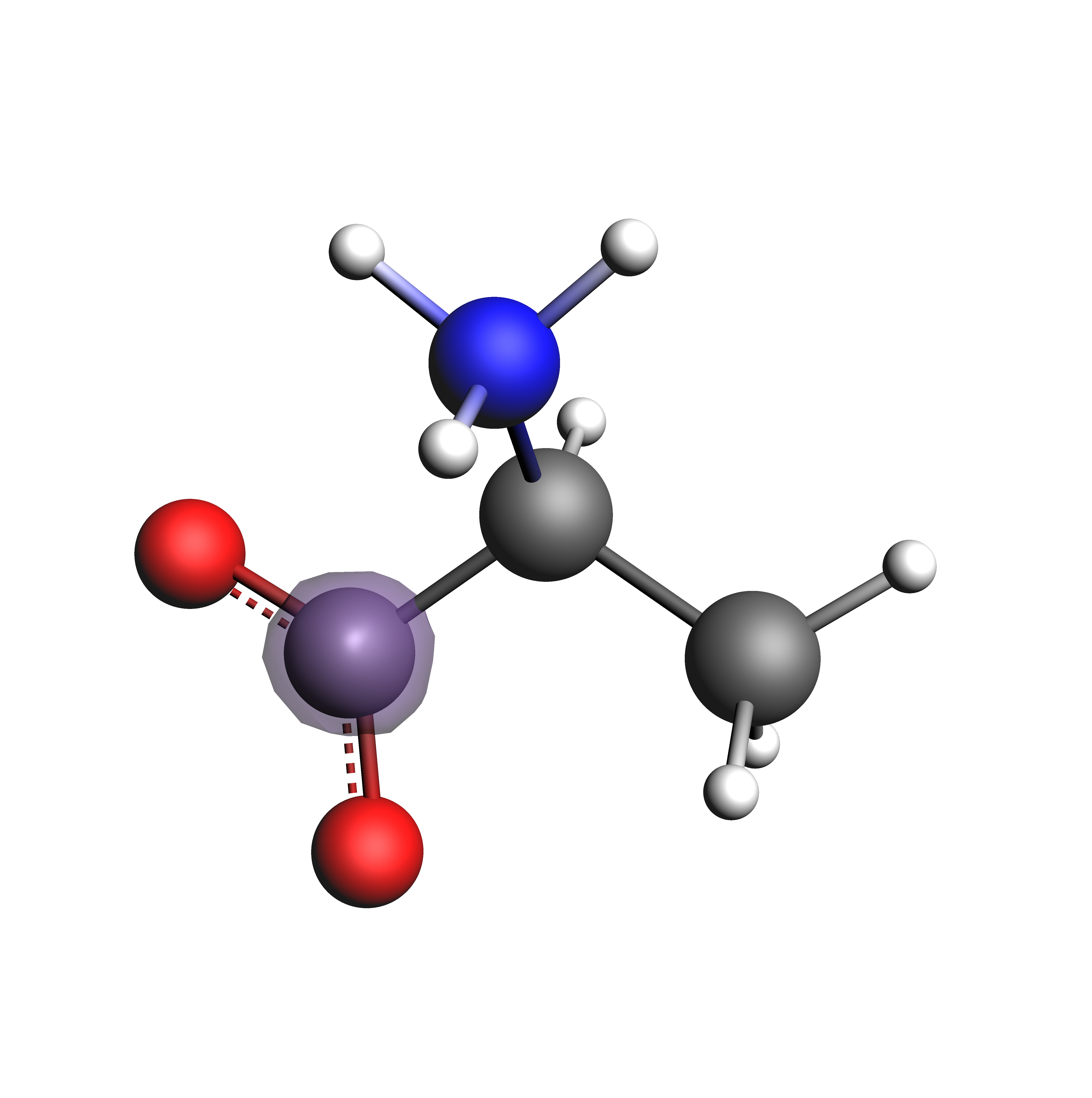}
    \includegraphics[width=0.30\linewidth]{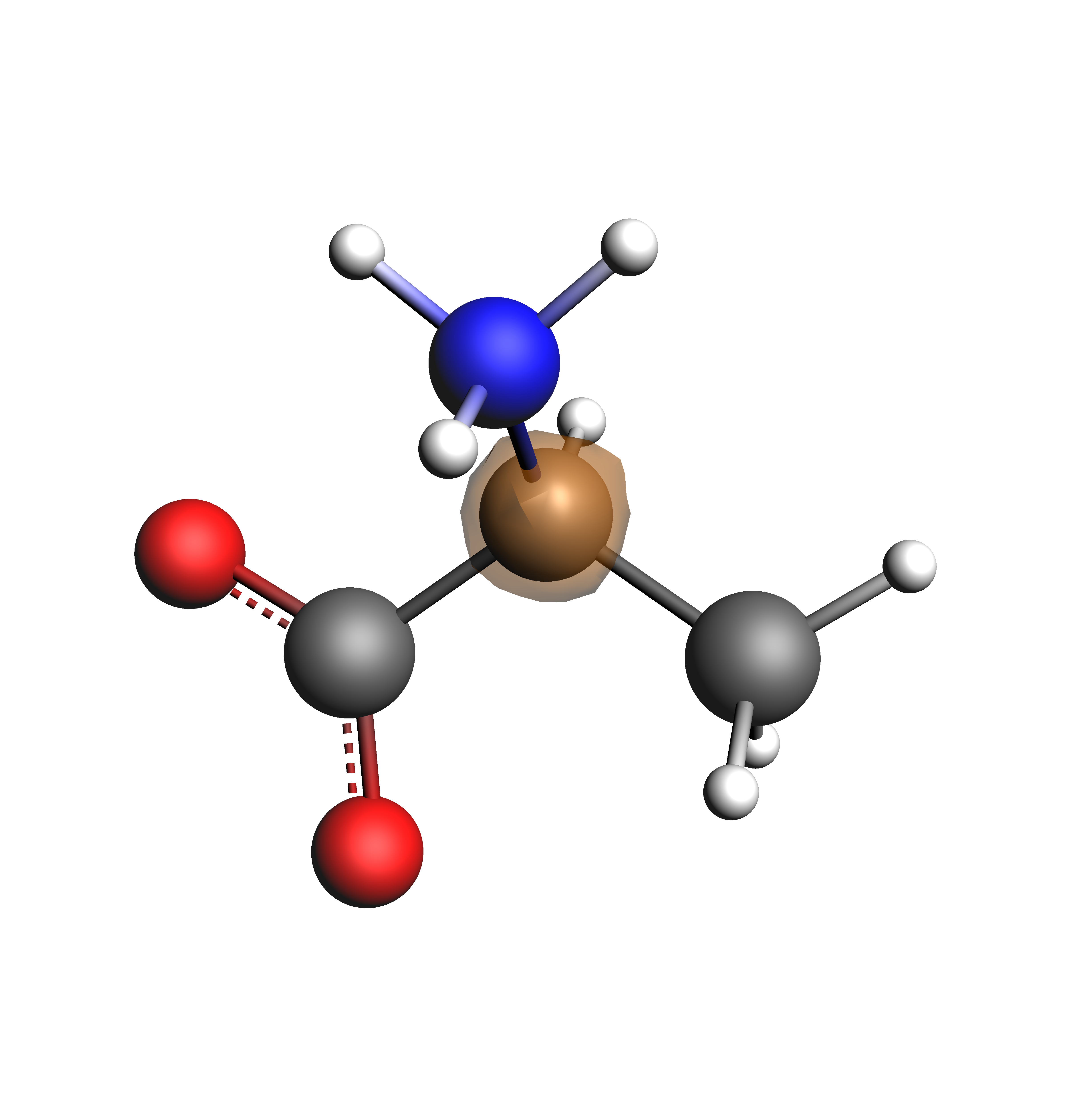}
    \includegraphics[width=0.30\linewidth]{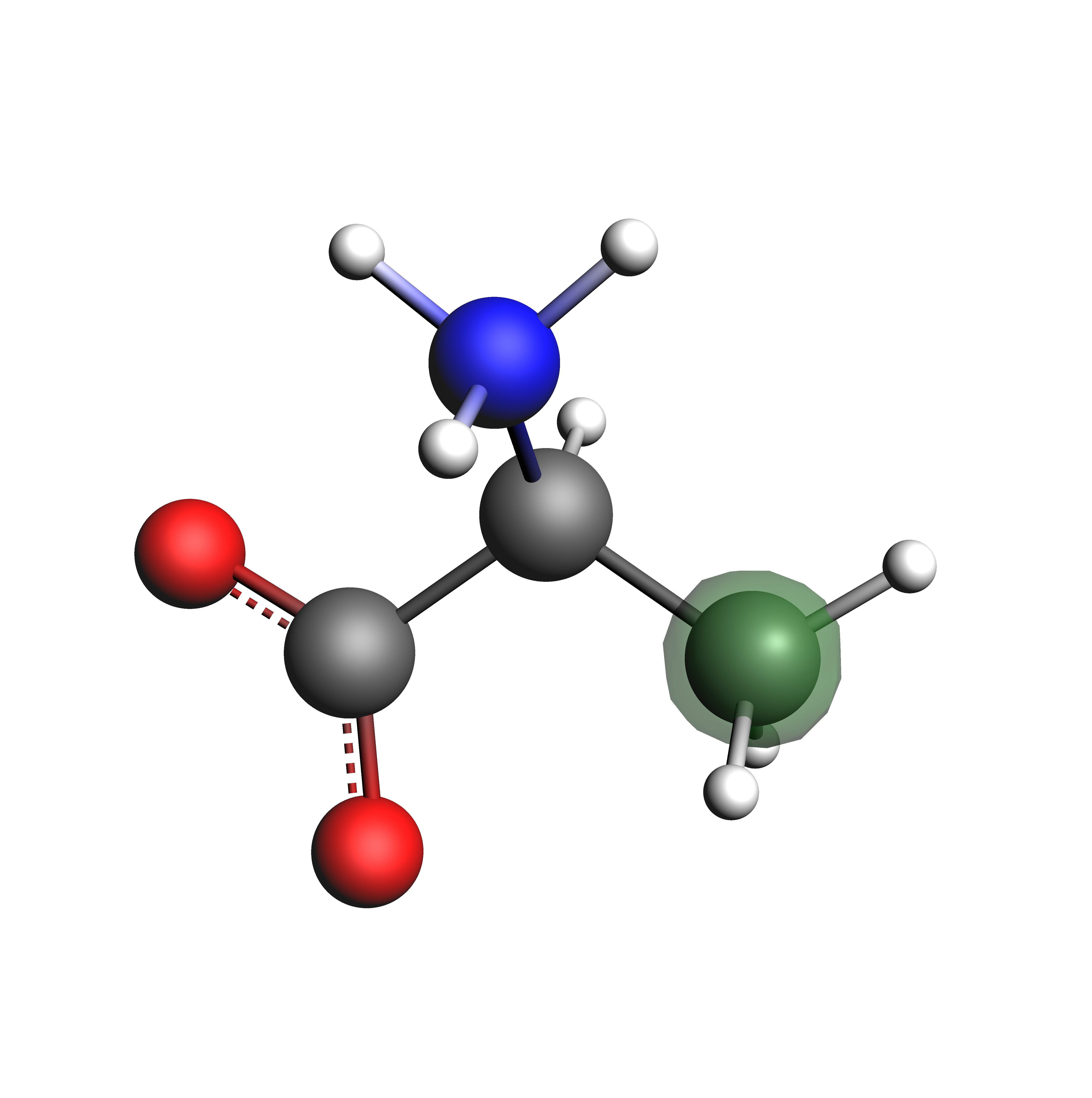}
    \caption{Spatial distributions of the three carbon C~$1s$ core
    orbitals of zwitterionic L-alanine. From left to right:
    carboxylate carbon C$_1$, $\alpha$ carbon C$_2$, and methyl carbon
    C$_3$.}
    \label{fig:carbon_orbitals}
\end{figure}

Zwitterionic alanine contains three chemically distinct carbon sites:
the carboxylate carbon C$_1$, the stereogenic $\alpha$ carbon C$_2$, and
the methyl carbon C$_3$ (see Fig. \ref{fig:carbon_orbitals}). This site resolution provides a direct test of
whether the environmental model correctly describes the markedly
different local electrostatic and hydrogen-bonding environments.

\begin{figure}[t]
    \centering
    \includegraphics[width=\linewidth]{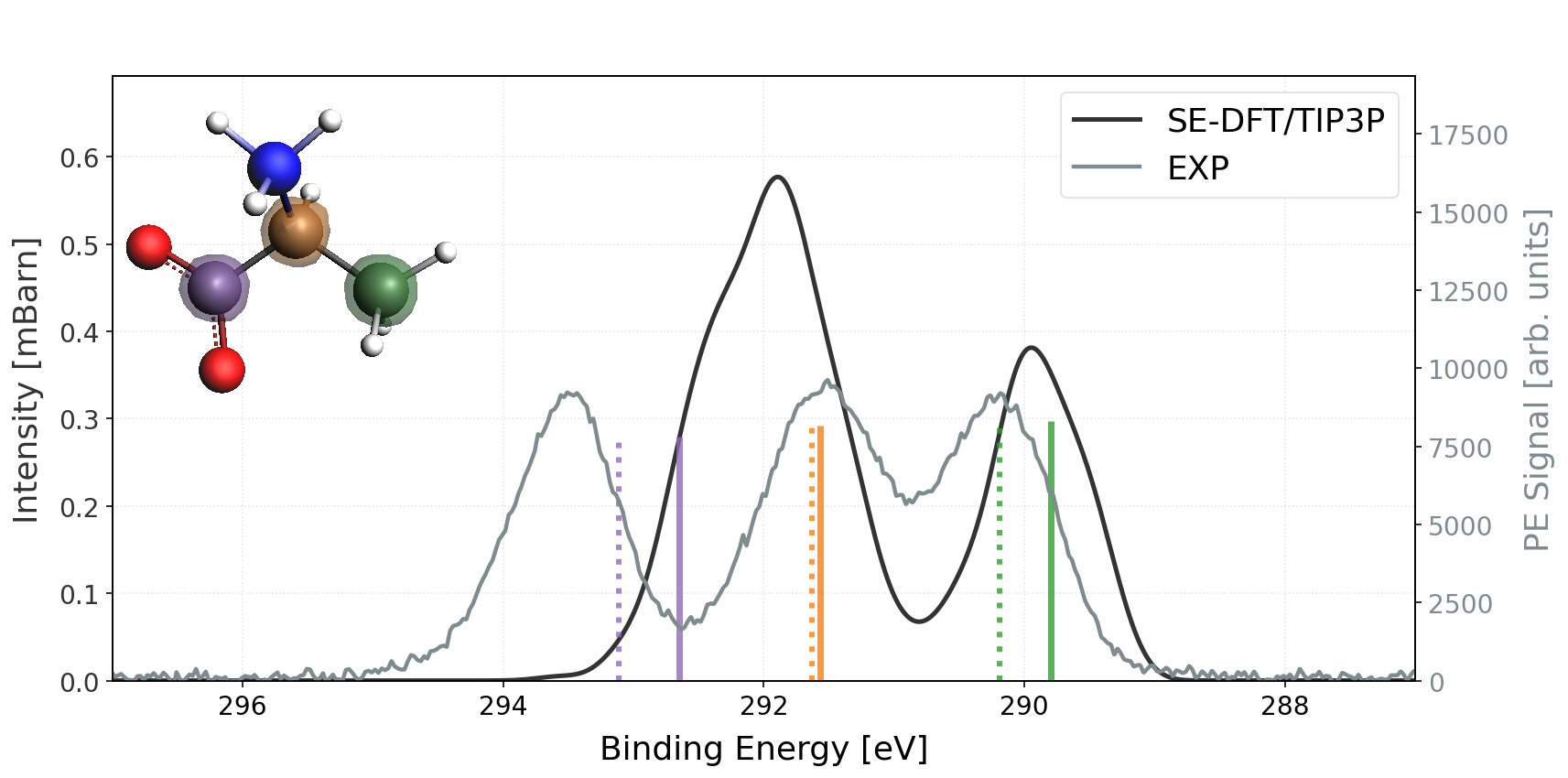}
    \caption{Calculated $\text{C}\,1\text{s}$ XPS photoelectron spectrum of zwitterionic L-alanine obtained using the non-polarizable SE-DFT/TIP3P embedding scheme, averaged over 200 spherical droplets. The individual core-level contributions are color-coded as follows: purple for the carboxylic carbon ($\text{C}_1$), orange for the $\alpha$-carbon ($\text{C}_2$), and green for the methyl carbon ($\text{C}_3$). Vertical colored bars represent the corresponding calculated binding energies. Dotted and solid vertical lines indicate the reference gas-phase and SE-DFT/COSMO values, respectively. Experimental spectra at $\text{pH}=6$, adapted from Stemer \textit{et al.} \cite{stemer2025photoelectron}, are displayed for comparison.}
    \label{fig:xps_tip3p}
\end{figure}

The non-polarizable electrostatic embedding scheme SE-DFT/TIP3P fails to resolve the individual contributions of the $\text{C}_1$ (carboxylic) and $\text{C}_2$ ($\alpha$-carbon) sites (see Fig~\ref{fig:xps_tip3p}) with respect to the experimental spectra at pH~6
reported by Stemer \textit{et al.}
\cite{stemer2025photoelectron}. The peaks corresponding to $\text{C}_1$ and $\text{C}_2$ are completely overlapped, resulting in a single, artificially broad convoluted band. While the position of the $\text{C}_2$ peak aligns reasonably well with the experimental evidence, the binding energy of the $\text{C}_1$ site is underestimated by approximately $1\,\text{eV}$. This discrepancy directly induces the unphysical overlap between the two signals. Also, a substantial overestimation of the spectral intensity is observed.

This behavior can be physically justified by noting that $\text{C}_1$ represents the carboxylic carbon atom; due to its highly polar nature and involvement in strong, localized hydrogen-bonding networks, the electronic structure of this site is exceptionally sensitive to explicit solvation and polarization effects, which a non-polarizable framework inherently fails to capture. Furthermore, within the non-polarizable approach, the $\text{C}_3$ (methyl) peak is left isolated, but it is shifted towards lower binding energies compared to the experimental reference.

A polarizable, yet implicit solvation description achieved via the SE-DFT/COSMO model, shows a persistent discrepancy of about $1\,\text{eV}$ between the carboxylic and $\alpha$-carbon peaks. Crucially, within this continuum framework, no configurational broadening is present, as the calculations rely on a single geometry optimized at the LB94/TZ2P-COSMO level, yielding a unique discrete eigenvalue for each core orbital. While the COSMO model successfully recovers the correct absolute position for the $\alpha$-carbon peak, it systematically underestimates the binding energies of the other two carbon sites. This is not surprising, since COSMO lacks any account of specific hydrogen-bonding effects, that are especially relevant for the carboxylic C, and for the methyl C, which is directly exposed to the solvent. Nevertheless, it is important to emphasize that by incorporating environmental polarization—even through a continuum dielectric approach—the relative peak splitting and overall electrostatic screening are better described than in non-polarizable TIP3P model, highlighting the dominant role of polarization over hydrogen bonding effects.

\begin{figure}[htbp]
    \centering
    \includegraphics[width=\linewidth]{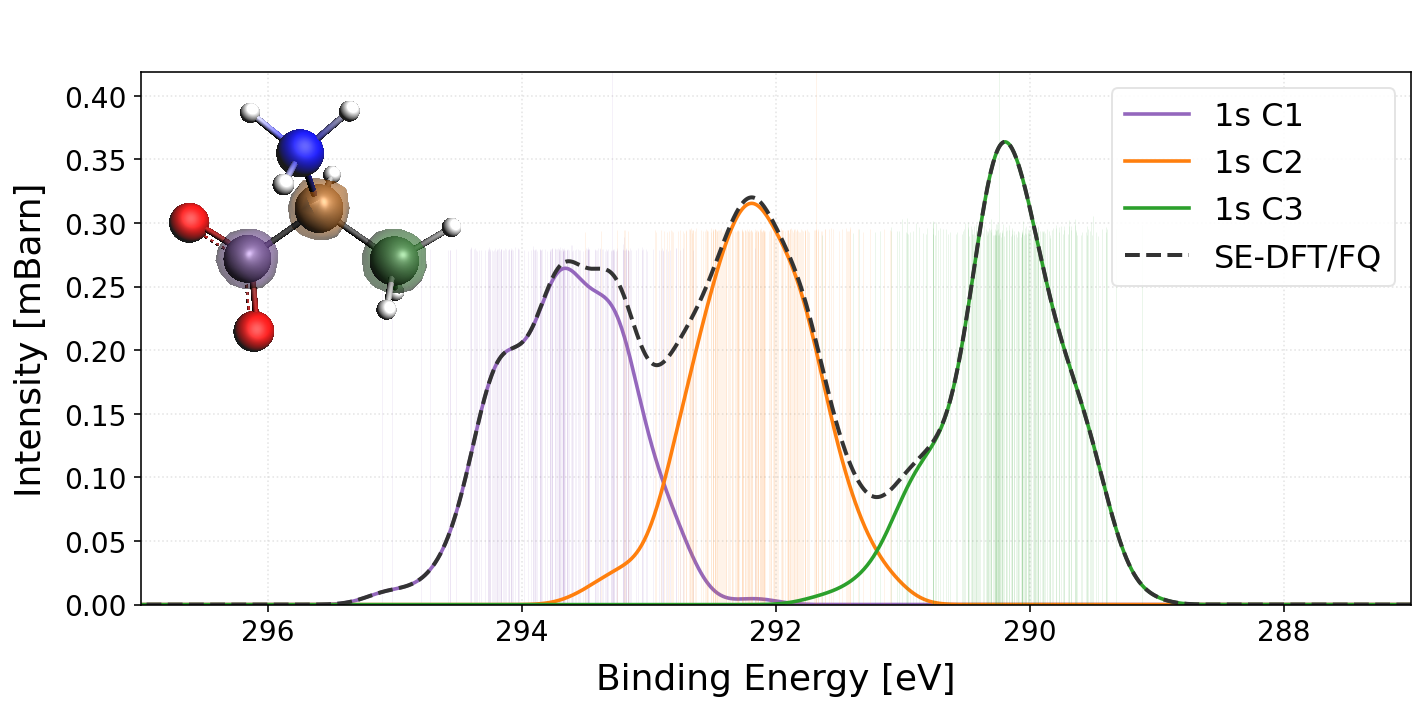}
    \caption{Computed core-level XPS photoelectron spectrum of zwitterionic L-alanine at SE-DFT/FQ level. The picture shows the averaged computed spectrum (dotted line) and the individual contributions for each carbon $1\text{s}$ core orbital, both in terms of raw data (sticks) and convolutions. }
    \label{fig:pes_giovannini_stick}
\end{figure}

We now move to the atomistic, fully polarizable SE-DFT/FQ approach. 
Figure~\ref{fig:pes_giovannini_stick} shows the raw data obtained from the calculations on the 200 droplets. Each stick represents the single calculation for the given droplet for the three carbon $1\text{s}$ core orbitals. Clearly, such a stick-spectrum already shows the spectral broadening which is observed in the final averaged envelope (dotted line). Such an inhomogeneous broadening directly reflects the sampling of the conformational phase space carried out via MD simulations.

\begin{figure}[t]
    \centering
    \includegraphics[width=\linewidth]{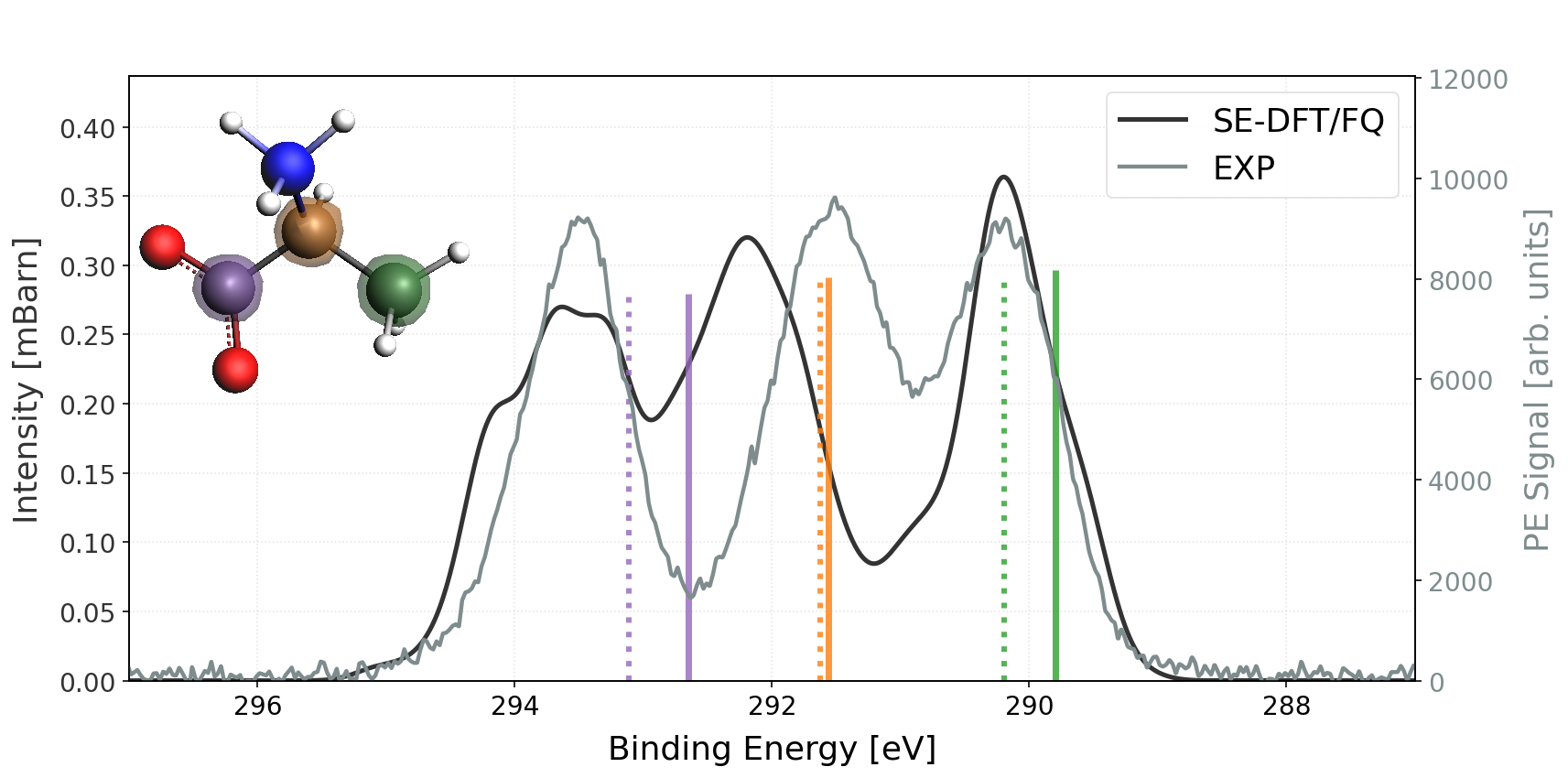}
    \caption{Calculated $\text{C}\,1\text{s}$ XPS photoelectron spectrum of zwitterionic L-alanine obtained using the polarizable SE-DFT/FQ embedding scheme, averaged over 200 spherical droplets. The individual core-level contributions are color-coded as follows: purple for the carboxylic carbon ($\text{C}_1$), orange for the $\alpha$-carbon ($\text{C}_2$), and green for the methyl carbon ($\text{C}_3$). Vertical colored bars represent the corresponding calculated binding energies. Dotted and solid vertical lines indicate the reference gas-phase and SE-DFT/COSMO values, respectively. Experimental spectra at $\text{pH}=6$, adapted from Stemer \textit{et al.} \cite{stemer2025photoelectron}, are displayed for comparison.} 
    \label{fig:xps_fqg}
\end{figure}

Fig.\ref{fig:xps_fqg} compares SE-DFT/FQ and experimental spectra. Gas-phase and SE-DFT/COSMO values are also reported for comparison. 

SE-DFT/FQ, which recovers both hydrogen bonding and polarization effects on the same footing, provides a substantially improved description
(\cref{fig:xps_fqg}) with respect to both SE-DFT/COSMO and SE-DFT/TIP3P (see Fig.\ref{fig:xps_tip3p}) not only in terms of spectral shifts, but also intensities.

The carboxylic carbon ($\text{C}_1$) peak is correctly positioned. This success stems from the fact that FQ describes the environment polarizability with sufficient accuracy to properly screen the carboxylic group. Consequently, the artificial overlap disappears, the experimental resolution between the $\text{C}_1$ and $\text{C}_2$ peaks is fully restored, and the overall convoluted spectrum matches the experimental features with high fidelity. Also, the correct alignment for the isolated $\text{C}_3$ peak is preserved. This comparative analysis demonstrates that the complete physical picture can only be recovered by simultaneously employing a polarizable and an atomistic description of the environment. Indeed, although core electrons are strongly localized near the nuclei, their binding energies are heavily modulated by the local electrostatic and polarization fields exerted by the immediate solvation shell.

It is also worth noting that the computed SE-DFT/FQ spectral profile strongly depends on the number of spherical droplets employed in the final average. A stable spectrum is recovered after 180 structures are averaged, while a small sampling set yields spectral profiles that do not match the experimental values, especially in terms of peaks' relative intensities. This behaviour is perfectly in line with previous studies of some of us. \cite{cappelli2016integrated, giovannini2020molecular}

\subsubsection{Photoelectron circular dichroism}

To start the discussion on PECD spectra, we take as reference the computed gas-phase profile of the dichroic asymmetry parameter 
$\beta_1$ within the energy range matching the experimental window. As it has been reported in the literature, the description of the PECD spectra for gas-phase molecules by SE-DFT approaches gives generally a fair account of the experimental results \cite{janssen2014detecting,nahon2015valence,dowek2022trends} 
The gas-phase spectrum exhibits a negative sign up to roughly $11.5\,\text{eV}$, switches to a positive sign with a low absolute intensity up to $13.5\,\text{eV}$, and reverts to negative values for the remainder of the analyzed interval. This behavior does not agree with the reported experimental trend in solution,\cite{stemer2025photoelectron}.

The quality of the gas-phase spectrum in reproducing experimental values is much lower than the implicit SE-DFT/COSMO, which recovers the negative sign between 11.59 and 12.60 eV, but gives a $\beta_1$ value nearly null above 13.62 eV.

\begin{figure}[t]
    \centering
    \includegraphics[width=\linewidth]{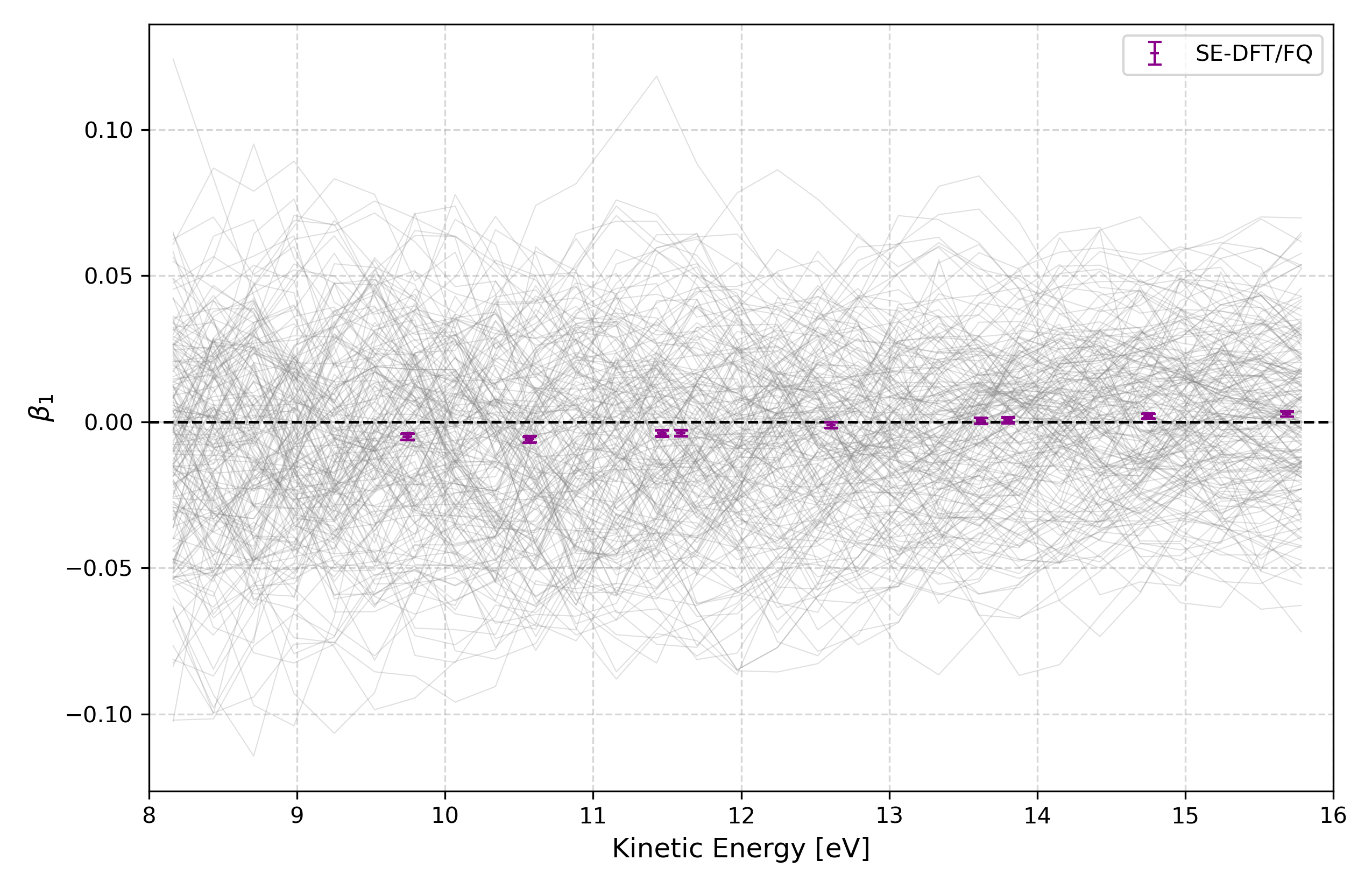}
    \caption{Computed SE-DFT/FQ C$_1$ PECD profiles.
     Thin gray curves represent calculations for a selection of 200 spherical droplets, while the violet squares indicate the average value (over 1000 droplets) at a given kinetic energy. 
     }
    \label{fig:pecd_snapshots}
\end{figure}

\cref{fig:pecd_snapshots} shows computed SE-DFT/FQ C$_1$ PECD profiles for a selection of spherical droplets, and the final value, which is obtained by averaging 1000 droplets.
The variability of $\beta_1$ as a function of the configuration is huge, both in sign, absolute values and overall profile. The mean $\beta_1$ value for a given kinetic energy is therefore not represented by a single preferred configuration, but originates from the average of the different profiles. Therefore, it is expected the final average $\beta_1$ value to arise from a subtle interplay of electronic, solvent electrostatics and polarization, and conformational effects. The observed behaviour is not surprising, and in line with previous studies of some of us, reporting strong configurational dependence of computed QM/FQ chiroptical properties.\cite{lipparini2013optical,egidi2019combined,giovannini2016effective,giovannini2017polarizable,giovannini2020theory}  

\begin{figure}[htbp]
    \centering
    \includegraphics[width=1\linewidth]{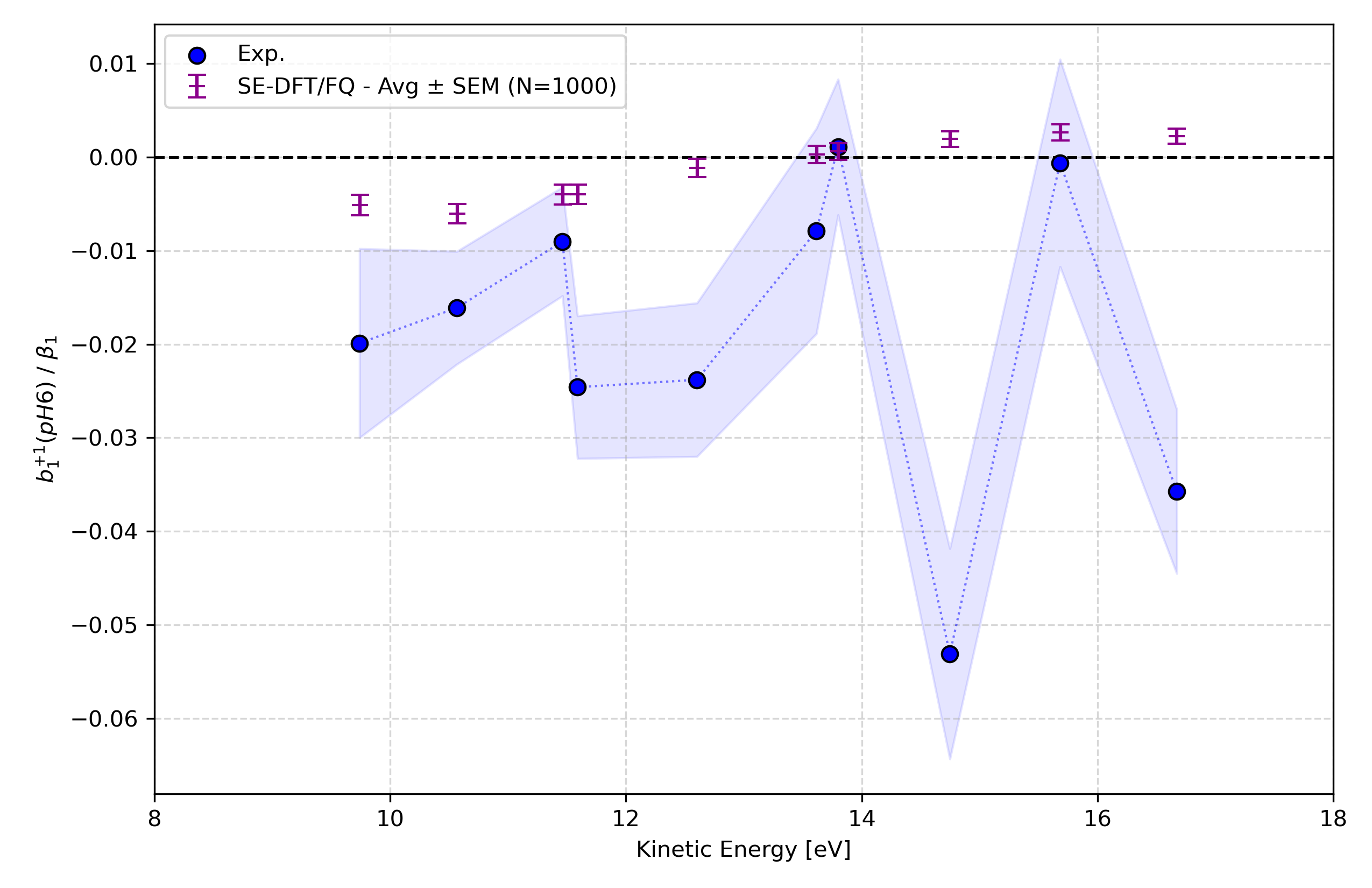}
    \caption{Calculated average SE-DFT/FQ $\beta_1$ for the carboxylic carbon C 1s photoionization of zwitterionic alanine. The standard error ($\pm\text{SEM}$) is reported. The average and SEM are obtained on the set of 1000 spherical droplets ensuring convergence in the average values. Experimental data, adapted from Ref.\cite{stemer2025photoelectron} are also shown for comparison.}
    \label{fig:pecd_c1s_carboxylic_sem}

\end{figure}

The average SE-DFT/FQ values (see also Fig.~\ref{fig:pecd_c1s_carboxylic_sem}) show a stable negative value up to approximately $13.5~\text{eV}$ , consistent with the sign of the experimental data.\cite{stemer2025photoelectron} A sign inversion (crossing zero into positive values) is observed at approximately $14~\text{eV}$. The experiments predict large negative values for the $\beta_1$ parameter at $14.75$ and $16.67~\text{eV}$, whereas our computational values stay small and positive (on average) in all the region above $14~\text{eV}$. At $15.68~\text{eV}$ the calculations match both sign and absolute value of the experimental value.

It is worth noting that the average $\beta_1$ values strongly depend on the number of structures employed in the averaging procedure. The dependence is substantially more dramatic than for XPS; the number of structures not only influences absolute intensities, but their sign, resulting in $\beta_1$ profiles that substantially differ until convergence is achieved. This pronounced dependence of chiral simulation in aqueous solution upon the sampling size confirms once again previous findings reported in the literature by some of us.\cite{lipparini2013optical,egidi2019combined,giovannini2016effective,giovannini2017polarizable,giovannini2020theory}

From the analysis of non-polarizable SE-DFT/TIP3P and SE-DFT/COSMO data, it emerges that both effects are important to align calculations with experimental data. 
Also, similarly to what is discussed above for XPS,  
SE-DFT/FQ data show a strong dependence on the employed FQ parametrization, being the least polarizable parametrization, FQ$_R$, in worst agreement with experiments.

Furthermore, in contrast to typical gas-phase PECD spectra, the values computed in solution show rather small and structureless behaviour. This attenuation of the overall PECD signal is due to both statistical ensemble averaging over solvent conformations and additional intermolecular scattering processes inherent to the liquid phase~\cite{hartweg2021condensation,pohl2022photoelectron}.
The absolute experimental magnitude could
therefore be attenuated relative to the calculated molecular response.

We finally note that the comparison reported above focuses primarily on the sign and broad energy dependence of the PECD profile rather than on point-by-point
agreement in magnitude. The present results are encouraging and generally consistent with the overall experimental pattern, however the calculation of PECD in solution is a very challenging task, and achieving a full evaluation and confidence in the quality of the theoretical results would require a much larger set of experimental and calculated data on a variety of systems.

\section{Summary, conclusions and future perspectives}

In this work, we have developed and validated an SE-DFT/FQ framework for the calculation of molecular photoionization observables in atomistic polarizable environments. The approach consistently transfers the mutually polarized ground-state QM/FQ description to the static-exchange Hamiltonian used to construct the continuum states, while retaining the computational efficiency required for ensemble calculations involving hundreds of solvent configurations.
Application of the method to the C~1s XPS spectrum and carboxylate-carbon PECD of aqueous L-alanine highlights the complementary and inseparable roles of a fully atomistic description of the system and environmental polarization. The non-polarizable TIP3P embedding fails to reproduce the experimental separation of the carbon core-level signals.
By contrast, the FQ polarizable embedding restores the main experimental XPS pattern and predicts a predominantly negative intrinsic PECD in the lower kinetic-energy region, in qualitative agreement with the available experimental data at pH~6. These results demonstrate that an atomistic representation of the solvent and its polarization response must be treated together to achieve a physically meaningful description of photoionization observables in aqueous solution.

The results also show that introducing environmental polarizability is not sufficient \emph{per se}. The quality of the calculated observables depends critically on the adopted FQ parametrization, since different descriptions of the magnitude and spatial distribution of the solvent response can produce qualitatively different continuum properties. A careful and independently validated parametrization of the environment is therefore an essential component of the proposed computational strategy.

Despite the very encouraging agreement obtained for aqueous alanine, the present results should not be interpreted as evidence that the same level of accuracy will necessarily be achieved for all molecular systems. The current investigation is restricted to a single prototypical solute, under experimental conditions corresponding to approximately pH~6. A broader validation and application study involving chemically diverse molecules in aqueous solution is therefore required to assess the generality, robustness, and predictive capability of the proposed approach. Remarkably, the methodology is not intrinsically restricted to aqueous environments and can also be extended to other solvents, provided that validated FQ parametrizations are available.\cite{ambrosetti2021quantum}

Moreover, the present embedding model describes solute-environment interactions exclusively in terms of electrostatics and mutual polarization. Non-electrostatic contributions, including short-range Pauli repulsion and dispersion,\cite{giovannini2017general} are not explicitly included. Although the present results suggest that electrostatic and polarization contributions capture the dominant environmental effects for the investigated system, non-electrostatic interactions may modify both the bound-state energetics and the scattering potential experienced by the outgoing electron, thereby affecting the final XPS and PECD profiles. Their incorporation into the theoretical framework is currently under development.

We finally note that the present calculations do not disentangle the contribution of individual solute-solvent hydrogen bonds from the overall response of the environment. Accordingly, although the results clearly establish the importance of an atomistic and polarizable description of the solvent, they do not yet permit a quantitative assignment of specific PECD features to individual hydrogen-bonding motifs. Such a decomposition would require a systematic classification of the configurational ensemble according to the hydrogen bonds formed around the relevant molecular sites, followed by a motif-resolved analysis of the corresponding photoionization observables. Also, any explicit, electronic, contribution of the closest water molecules is not included in our model. This kind of analyses lies beyond the scope of the present work but represents an important direction for future investigations.

Ongoing studies are extending the approach to the different protonation states of alanine populated under different pH conditions, as well as to more complex molecular systems. Particular attention is being devoted to flexible molecules displaying significant conformational freedom. In such cases, an additional level of complexity is introduced and requires statistically converged sampling of both molecular and solvent degrees of freedom.

\section*{Acknowledgments}
This work was supported by the European Union's Horizon Europe Research and Innovation Programme under Grant Agreement  101169312. The authors gratefully acknowledge the Center for High-Performance Computing (CHPC) at the Scuola Normale Superiore for providing access to the computational infrastructure used in this work.
The authors gratefully acknowledge Prof.~Bernd Winter and Dr.~Dominik Stemer for valuable discussions and for providing the raw experimental data used to plot the spectra shown in the figures.

%
\section*{Author Declarations}

\subsection*{Conflict of Interest}

The authors have no conflicts to disclose.

\subsection*{Author Contributions}

Giovanni Nottoli: Software, Validation, Formal analysis,
Investigation, Visualization, Writing--original draft.
Piero Decleva: Software, Supervision, Writing--review and
editing.
Chiara Cappelli: Conceptualization, Methodology, Supervision, Project
administration, Funding acquisition, Writing--review and editing.

\section*{Data Availability}

The data that support the findings of this study are available from the
corresponding author upon reasonable request.

\bibliographystyle{aipnum4-2}
\bibliography{biblio_def_abbrev,references}

\end{document}